\documentstyle[12pt]{article}
\def\Re{{\cal R \mskip-4mu \lower.1ex \hbox{\it e}\,}}
\def\Im{{\cal I \mskip-5mu \lower.1ex \hbox{\it m}\,}}

\def\sub#1{_{\lower.25ex\hbox{$\scriptstyle#1$}}}
\def\sul#1{_{\kern-.1em#1}}
\def\sll#1{_{\kern-.2em#1}}
\def\sbl#1{_{\kern-.1em\lower.25ex\hbox{$\scriptstyle#1$}}}
\def\ssb#1{_{\lower.25ex\hbox{$\scriptscriptstyle#1$}}}
\def\sbb#1{_{\lower.4ex\hbox{$\scriptstyle#1$}}}

\def\gev{\,{\rm GeV}}

\def\mh{\ifmmode m\sbl H \else $m\sbl H$\fi}
\def\mch{\ifmmode m_{H^\pm} \else $m_{H^\pm}$\fi}
\def\mt{\ifmmode m_t\else $m_t$\fi}
\def\mc{\ifmmode m_c\else $m_c$\fi}
\def\mz{\ifmmode M_Z\else $M_Z$\fi}
\def\mw{\ifmmode M_W\else $M_W$\fi}
\def\mws{\ifmmode M_W^2 \else $M_W^2$\fi}
\def\mhs{\ifmmode m_H^2 \else $m_H^2$\fi}
\def\mzs{\ifmmode M_Z^2 \else $M_Z^2$\fi}
\def\mts{\ifmmode m_t^2 \else $m_t^2$\fi}
\def\mcs{\ifmmode m_c^2 \else $m_c^2$\fi}
\def\mchs{\ifmmode m_{H^\pm}^2 \else $m_{H^\pm}^2$\fi}
\def\ztwo{\ifmmode Z_2\else $Z_2$\fi}
\def\zone{\ifmmode Z_1\else $Z_1$\fi}
\def\mtwo{\ifmmode M_2\else $M_2$\fi}
\def\mone{\ifmmode M_1\else $M_1$\fi}
\def\tb{\ifmmode \tan\beta \else $\tan\beta$\fi}
\def\xw{\ifmmode x\sub w\else $x\sub w$\fi}
\def\ch{\ifmmode H^\pm \else $H^\pm$\fi}
\def\lum{\ifmmode {\cal L}\else ${\cal L}$\fi}
\def\inpb{\ifmmode {\rm pb}^{-1}\else ${\rm pb}^{-1}$\fi}
\def\infb{\ifmmode {\rm fb}^{-1}\else ${\rm fb}^{-1}$\fi}
\def\epem{\ifmmode e^+e^-\else $e^+e^-$\fi}
\def\ppb{\ifmmode \bar pp\else $\bar pp$\fi}

\def\half{\textstyle{{1\over 2}}}

\newskip\zatskip \zatskip=0pt plus0pt minus0pt
\def\matth{\mathsurround=0pt}

\def\atversim#1#2{\lower0.7ex\vbox{\baselineskip\zatskip\lineskip\zatskip
  \lineskiplimit 0pt\ialign{$\matth#1\hfil##\hfil$\crcr#2\crcr\sim\crcr}}}

\renewcommand{\thefootnote}{\fnsymbol{footnote}}

\begin{document} \begin{titlepage}
\setcounter{page}{1}
\thispagestyle{empty}
\rightline{\vbox{\halign{&#\hfil\cr
&ANL-HEP-PR-92-37\cr
&May 1992\cr}}}
\vspace{0.2in}
\begin{center}

{\Large\bf
New Probes for Extended Gauge Structures at HERA}
\footnote{Research supported by the
U.S. Department of
Energy, Division of High Energy Physics, Contracts W-31-109-ENG-38
and W-7405-Eng-82.}
\medskip

\normalsize THOMAS G. RIZZO
\\ \smallskip
High Energy Physics Division\\Argonne National
Laboratory\\Argonne, IL 60439\\
\smallskip
and\\
\smallskip
Ames Laboratory and Department of Physics\\
Iowa State University\\ Ames , IA 50011\\

\end{center}

\begin{abstract}

Doncheski and Hewett have recently shown that the ratio of neutral current to
charged current cross sections, $R={\sigma_{NC}}/{\sigma_{CC}}$, can provide a
more sensitive probe for the existence of heavy leptoquarks at HERA than
the usual proceedure which makes use of neutral current asymmetries.
The apparent reason for this is that the
Standard Model expectations for {\it both} of these cross sections are modified
by the existence of such particles in a semi-coherent manner. In this paper we
apply this technique to extended electroweak models whose spectrum contains
both a $W'$ and a $Z'$. We find that measurements of $R$ can, for some
models, substantially increase the HERA search range for new gauge bosons
beyond that which can be probed using the more conventional asymmetries.

\end{abstract}



\renewcommand{\thefootnote}{\arabic{footnote}} \end{titlepage}


The start-up of the HERA ep collider opens a new regime in which to explore
for
physics beyond the Standard Model(SM){\cite {bigref}}. As such, it is
important to be able to extend the search range for potential new physics as
much as
possible given the limitations of luminosity and center of mass energy. For
example, previous to
the recent work of Doncheski and Hewett(DH){\cite {DH}}, it had been thought
that
HERA could search for the leptoquarks arising in $E_6$ models{\cite{HR}} up to
masses comparable to the machine's center of mass energy, ${\sqrt s}$=314 GeV,
even for relatively weak leptoquark coupling strengths in comparison to
electromagnetism. Such searches could be performed
either by direct production or by hunting for deviations from the SM
predictions for the values of various
neutral current asymmetries{\cite {OLQ}}. DH have, however, shown that it will
possible to look for still heavier leptoquarks (with masses even as large as
800 GeV) provided their coupling stength is not too small relative to
electromagnetism. The key to their analysis was to notice that the SM
prediction
for the ratio of
neutral to charged current cross sections for unpolarized beams,
$R={\sigma_{NC}}/{\sigma_{CC}}$, is modified in a semi-coherent manner by the
existence of leptoquarks and that various systematic errors, such as those due
to luminosity and structure function uncertainties, mostly cancel in such a
ratio.

The purpose of this paper is to explore whether other sorts of new physics, in
particular, models with extended gauge sectors, can be more sensitively probed
using the ratio $R$. While the possibility of using HERA to search for a new
gauge boson, $W'$ or $Z'$, has been widely discussed in the literature
{\cite {bigref2}}, such analyses have failed to examine the coherent influence
of these two particles $simultaneously$ in models where both are present,
hence making use of the DH technique.
Thus in this paper we will seek to
explore whether the ratio $R$ can extend the previously obtained search ranges
for new gauge bosons at HERA. We will then compare these new limits with what
may be
obtainable via direct production searches at the Tevatron. We will find that
at least for some extended models, the $95\%$ CL search limits obtainable
$indirectly$ from HERA are more than comparable to those arising from the
parallel direct searches at the Tevatron. As we will see, an important
ingredient in this analysis is a relationship between the $W'$ and $Z'$
masses.

Since $R$ will clearly be most sensitive to the existence of new gauge bosons
when both a $W'$ {\it and} a $Z'$ are present, we will restrict our attention
to extended gauge models where both kinds of particles are predicted to exist.
(We have checked that in models
with only a $Z'$ $or$ a $W'$ that the mass limits obtainable from the ratio
$R$ are comparable or
inferior to the more standard results obtained via the examination of various
polarization
asymmetries as one would naively expect.)  To be specific, we will restrict
our attention to the three models which
follow: ($i$) the Left-Right Symmetric Model(LRM){\cite{LRM}},
wherein the $W'$ couples to right-handed currents and the only free parameters
(other than the $Z'$ mass)
are the ratio of right-handed to left-handed gauge couplings,
$\kappa = g_R/g_L$, and the structure of the $SU(2)_R$-breaking scalar sector
as expressed through the $W'$ and $Z'$ mass relationship
\medskip
\begin {equation}
{M_{W_2}^2\over {M_{Z_2}^2}} =  {{(1-x_w)\kappa^2-x_w}\over {\rho_R(1-x_w)
\kappa^2}}
\end {equation}
\medskip
where $M_{W_2}$ and $M_{Z_2}$ are the $W'$ and $Z'$ masses respectively,
$x_w = sin^2\theta_w$, and $\rho_R$ takes on the value 1(2) if the $SU(2)_R$
breaking sector consists of Higgs doublets(triplets). (In this equation, and
in the various gauge model couplings we will assume for numerical purposes that
$x_w$=0.2325, which is its
effective value at the weak scale{\cite {LEP}}.) In calculating matrix
elements, we will assume that the flavor-mixing matrix for the right-handed
currents has essentially the same structure as the conventional
Kobayashi-Maskawa matrix for left-handed currents in the sense that it is
nearly diagonal. (We remind the cautious
reader that this need not be the case.) $(ii)$ the 'Un-unified' Model of
Georgi, Jenkins, and Simmons(GJS)
{\cite {HARV}} in which
the quarks and leptons couple to {\it different} $SU(2)$ gauge sectors. The
two new
gauge bosons in this model are purely left-handed and degenerate in mass to a
very high level of accuracy.
Their couplings depend only upon a single mixing-angle parameter,
$0.22<s_\phi<0.99$. $(iii)$ the model of Bagneid, Kuo, and Nakagawa(BKN)
{\cite {KUO}}, in which the $W'$ and $Z'$ are essentially degenerate, as in
the GJS case,
but couple differently to the third generation than the first two. This model
has no additional free parameters and both the $W'$ and $Z'$ are purely left-
handed.

Although this is not an exhaustive list of models it is fairly
representative of those existing in the literature; for detailed descriptions
of these models we refer the interested reader to the original references.
Some of these other models, which we will not discuss here, are
clearly distinguishable from the SM since they predict that the exchange of
a $W'$ will lead to new particle production at the leptonic and/or hadronic
vertex; see for example{\cite {ALRM}}.

To calculate the ratio $R$ we first note that both  $\sigma_{NC}$ and
$\sigma_{CC}$ appearing in the definition of $R$ are $unpolarized$ cross
sections and we will
assume an equal incoming flux of both $e^+$ and $e^-$ beams, i.e., the cross
sections are charge averaged.
In order to seperate neutral current from charged current events, we employ
the same kinematic cuts as DH; to remove the major part of the photon pole and
to advoid the region where structure function uncertainties are largest we
make the restriction $0.1\le x\le 1$. Given a fixed value of x we then
further restict the variable $y$ to the range
\medskip
\begin{equation}
{\rm  max}(0.1, y_{min})\le y\le {\rm min}(1, y_{max}) \,,
\end{equation}
\medskip
where $y_{max,min}$ are defined in terms of either a $p_T$ cut on the outgoing
electron in the neutral current case or a $missing$ $p_T$ cut, due to the
outgoing neutrino, in the charged current case:
\medskip
\begin{equation}
y_{max,min}=\half \left[ 1 \pm \sqrt{1 - { \mbox{4($p_T^{cut})^2$} \over
\mbox{$xs$}}} \;\; \right] \,.
\end{equation}
\medskip
This cut not only helps us to seperate the events into the NC and CC catagories
 but also helps to increase the influence of the new heavy particles we have
introduced (be they new gauge bosons or leptoquarks) and further reduces the
relative contribution of the photon pole.
DH make use of the following specific $p_T$ cuts: $p_T(e)>5\gev$ for neutral
currents
and $\not p_T(\nu) > 20\gev$ for charged currents. We employ the Harriman
et al.,
HMRS-B{\cite{HMRS}}
parton distributions as our default in performing our calculations but we
have checked that other distributions, such as those of Morfin and Tung
{\cite{MT}} do not lead to results different than those quoted below by more
than $5\%$. The equations for the various differential cross sections we need
to evaluate have been given completely elsewhere{\cite{DH,HR,bigref2}} and so
will not be given explicitly here. These
various cross sections are first calculated within the SM in order to obtain
the ratio $R$ and then again within the context of various extended models for
different values of the model parameters and as a function of the $Z'$ mass,
$M_{Z_2}$. (The corresponding $W'$ mass is then given in terms of the model-
dependent mass
relationships discussed above.) The value of $R$ obtained within the extended
model is then compared with SM expectations via a $\chi^2$ analysis. Since
most of the systematic errors in $R$ are expected to cancel in the taking the
ratio of cross sections, the dominant error in $R$ will be purely statistical
and easily calculated for a fixed integrated luminosity, ${\cal L}$:
\medskip
\begin{equation}
{\delta R\over R}={1\over\sqrt{N_{NC}}}\oplus{1\over\sqrt{N_{CC}}} \,,
\end{equation}
\medskip
with $N_{NC}$ and $N_{CC}$ simply given by $N_{NC,CC}={\cal L}\sigma_{NC,CC}$.
The `$\oplus$' in the above equation implies that the errors are to be added
in quadrature. The value of the $Z'$ mass is then raised from some small value
until the deviation from
the expectations of the SM reach the $95\%$ CL; this particular value of
$M_{Z_2}$ (and correspondingly $M_{W_2}$) is then the search limit which we
quote below.

Fig.1 shows the limits that we obtain by this proceedure for the various
models discussed above as functions of the HERA integrated luminosity per
$e^{\pm}$ beam. We first note that although the couplings of the GJS model
depend on the parameter $s_{\phi}$, the limits we obtain are independent of
this parameter. The reason for this is that all factors of $s_{\phi}$ cancel in
the
product of couplings which appear in the matrix element $and$ the $s_{\phi}$
dependence of the $Z'$ and $W'$ widths, appearing in the corresponding
propagators, is relatively unimportant as these particles are exchanged in
the t-channel. We also note that the limits we obtain from the ratio $R$ are
highly model dependent. In the case of the LRM, we explicitly display the
results where the $SU(2)_R$ breaking sector consists only of scalar doublets.
The corresponding limits for breaking via isotriplets are smaller and can be
obtained approximately by simply scaling the isodoublet results by a factor of
0.85.

How do these results compare to the corresponding
limits which can be obtained from asymmetry measurements? A recent analysis
from the Snowmass 1990 Summer Study for these same models{\cite {bigref2}}
that
assumed $80 \%$ beam polarization and an integrated luminosity of 400
$pb^{-1}$ (distributed equally among the $e^{\pm}_{L,R}$ beams) obtained the
following limits on $M_{Z_2}$: 520 GeV for the GJS model, 380 GeV for the LRM
with $\kappa$=1,
and 350 GeV for the model of BKN. (The limits in the GJS model were also found
to be $s_{\phi}$ independent in this case for the same reasons as above. Of
course, no assumption about the interreltionship of the $W'$ and $Z'$ masses
was necessary to obtain these asymmetry results.) Comparing with Fig.1, we
see that the
limits obtainable from $R$ $without$ beam polarization and equivalent total
integrated luminosity are substantially larger in both the GJS and BKN cases
than what is obtainable using asymmetries but
only comparable limits are found in the LRM case assuming $\kappa$=1
and isodoublet breaking of $SU(2)_R$. Thus we see, at least for the models we
have examined, that the ratio $R$ does at least as well, and in most cases
better, than asymmetries in probing the extended gauge sector $provided$ we
can input some relationship between the $W'$ and $Z'$ masses.

The improved limits on new gauge bosons obtainable from $R$ now allow HERA
to be competative with, and in some cases superior to, the Tevatron in probing
extended gauge sectors. This is illustrated in Fig.2 which shows the search
limits for the GJS model at the Tevatron as a function of $s_{\phi}$ for
various integrated luminosities. (In obtaining these limits we have used the
electron and muon efficiencies as reported by the CDF collaboration
{\cite {CDF}} and we reproduce their quoted search limits for the new gauge
bosons of other extended electroweak gauge models
within the errors associated with structure function uncertainties.)
Unlike the situation at HERA, the Tevatron limits
are seen to be relatively sensitive to the value of $s_{\phi}$ even though the
$s_{\phi}$ dependence cancels in the product of quark and lepton couplings as
it does in ep collisions.
The reason for this is that the production rate, e.g., for lepton pairs, is
also quite
sensitive to the $s_{\phi}$ dependence of the $Z'$ width as the $Z'$ is now
exchanged in the s-channel and appears as a resonance in the parton level
subprocess. (The same is true for the production of a lepton plus neutrino
final state in
the case of $W'$). For  values of $s_{\phi}$, near the extrema of the allowed
range, the widths of the
$Z'$ and $W'$ become quite large thus supressing the leptonic production
cross section.
The actual limits we show in Fig.2 are those which arise from $W'$ production
as the cross section times branching ratio is about an order of magnitude
larger in this case than the corresponding one for the $Z'$. We then can simply
use the fact that
$M_{Z_2}$ = $M_{W_2}$ for the GJS model to quote a limit on the $Z'$ and
compare with the corresponding results obtained for HERA. A last caveat for
the case of the Tevatron is the assumption that the $Z'$ and $W'$ only decay
into SM particles when calculating total widths. As discussed above, the limits
we obtain at the Tevatron are quite sensistive to the $Z'$ and/or $W'$
widths. If
additional decay modes are available for the $Z'$ and/or $W'$ the
cross section times branching ratio will be reduced resulting in a weaking of
the limits we show in the figures. Thus, for each model, we only show the
$best$ that can be done at the Tevatron for a fixed integrated luminosity.
In the case of the GJS model, this leads us to conclude that HERA is a better
probe of the extended gauge sector than the Tevatron even if the HERA
integrated luminosity per beam is substantially smaller. For example, with
only a modest 50 $pb^{-1}$/$e^{\pm}$ beam at HERA, we can obtain a limit on
the $Z'$ mass in the GJS model of 640 GeV {\it independently} of the value of
$s_{\phi}$. To cover this same range of parameters at the Tevatron would
require integrated luminosities in the neighborhood of 100 $pb^{-1}$ or higher.
 As integrated luminsities increase at both machines, the HERA limits pull far
ahead of those obtainable at the Tevatron. Of course, to truely make a
comparison we would need to know the time evolution of the integrated
luminosity at both colliders.

The situation is less clear for the other two models. In the BKN case, since
there are no additional free parameters, we show in Fig.3 the search limit
for a $Z'$ or $W'$ as a function of the Tevatron integrated luminosity for
the same set of assumptions as in the GJS model case discussed above. Note
that the search limit rises almost linearly with the log of the integrated
luminosity. For example, assuming an
integrated luminosity of 50 $pb^{-1}$ per beam at HERA the search limit we
obtain from
Fig.1 is 430 GeV. To reach the same limit at the Tevatron would only require
an integrated luminosity of 7 $pb^{-1}$, not far from the present value. In
general, we find that the Tevatron and HERA do comparably well for this model
provided the $Z'$ and $W'$ do $not$ have additional decay modes
which would contribute substantially to their total widths. If such modes $do$
exist,
then HERA will provide the stronger limit for the BKN model case as well as
the GJS case.

The situation for the LRM is a bit more complex since the relationship between
the $W'$ and $Z'$ masses is no longer so trivial as in either the GJS or BKN
models. Just as in either of these scenarios, however, the Tevatron
limit on
the $W'$ mass will be substantially stronger than the corresponding one for
the $Z'$. Fig.4a shows the explicit limit on $M_{W_2}$ as a function of
$\kappa$ for different integrated luminosities at the Tevatron. Assuming
either the isodoublet
or isotriplet mechanism for $SU(2)_R$ breaking, these limits on the $W'$ can
be converted to ones on the $Z'$ which are shown in Figs.4b and 4c. Fig.4d
explicitly shows the relationship between the $Z'$ and $W'$ masses in the LRM
for both $SU(2)_R$ breaking scenarios needed to obtain Figs.4b and 4c from
Fig.4a. If we search
for the $Z'$ in this model $directly$ via the lepton pair signature at the
Tevatron, we would
instead obtain the result shown in Fig.4e which assumes for simplicity that
$\kappa$=1. We see that even in the case where $SU(2)_R$ breaking occurs via
triplets, the indirect limits on the $Z'$ using the $W'$ data and the mass
relationships is at least as strong as the direct $Z'$ search limit. Comparing
with Fig.1 for HERA, we observe that for the LRM case the Tevatron limit on
the $Z'$ mass is always as good or better than what is obtainable at HERA
using the ratio $R$ for either of the two $SU(2)_R$ symmetry breaking
scenarios.

In this paper we have attempted to extend the search limits for new gauge
bosons at HERA by using the ratio $R$ introduced by Doncheski and Hewett to
search for leptoquarks with masses in excess of the HERA center of mass energy.
 The main results of this analysis are as follows:

$(i)$ For the GJS model, the HERA limits were substantially improved by using
the ratio $R$ in comparison to the usual asymmetry technique and were found
to be independent of the parameter $s_{\phi}$. For this model, HERA was shown
to provide a stronger constraint on the $Z'$ mass than the Tevatron.

$(ii)$ In the case of the BKN model, HERA limits were somewhat improved via
the $R$ ratio so that the Tevatron and HERA limits were now found to be
roughly comparable in their abilities to explore for new gauge bosons. The
Tevatron limits would prove inferior to those obtained from HERA $if$ the
new gauge bosons were to decay substantially into non-SM final states.

$(iii)$ For the LRM case, the HERA limits were not substantially altered by
making use of $R$. The $indirect$ Tevatron limits on the $Z'$ mass which
followed from the $W'$ and $Z'$ mass relationship were always superior to
those obtainable at HERA. The $direct$ $Z'$ search limits at the Tevatron were
$also$ shown to be superior to what is obtainable at HERA. This situation
$might$ be substantially modified if non-SM final states resulted in
significant changes
in the expectations for the $Z'$ and $W'$ total widths.

$(iv)$ Clearly the ratio $R$ provides a useful tool in probing for extended
gauge sectors at HERA when both a $W'$ and a $Z'$ are present, doing as well
as or better than neutral current polarization asymmetries in all the cases
we have examined.  Of course, to employ the $R$ ratio technique to search for
new gauge bosons the models we examine $must$ predict a relationship between
the $Z'$ and $W'$ masses.

Perhaps such signatures at HERA will provide the first evidence for new
physics beyond the Standard Model.

\medskip
\vskip.75in
\centerline{ACKNOWLEDGEMENTS}

The author would like to thank JoAnne Hewett and Mike Doncheski for
discussions related to this
work and Wesley Smith for his continual encouragement to explore the physics
at HERA.
This research has been supported supported in part by the U.S.~Department
of Energy under contracts W-31-109-ENG-38 and W-7405-ENG-82.

\newpage

%
\def\MPL #1 #2 #3 {Mod.~Phys.~Lett.~{\bf#1},\ #2 (#3)}
\def\NPB #1 #2 #3 {Nucl.~Phys.~{\bf#1},\ #2 (#3)}
\def\PLB #1 #2 #3 {Phys.~Lett.~{\bf#1},\ #2 (#3)}
\def\PR #1 #2 #3 {Phys.~Rep.~{\bf#1},\ #2 (#3)}
\def\PRD #1 #2 #3 {Phys.~Rev.~{\bf#1},\ #2 (#3)}
\def\PRL #1 #2 #3 {Phys. Rev. Lett.~{\bf#1},\ #2 (#3)}
\def\RMP #1 #2 #3 {Rev.~Mod.~Phys.~{\bf#1},\ #2 (#3)}
\def\ZP #1 #2 #3 {Z.~Phys.~{\bf#1},\ #2 (#3)}
\def\IJMP #1 #2 #3 {Int.~J.~Mod.~Phys.~{\bf#1},\ #2 (#3)}

\newpage

%
{\bf Figure Captions}
\begin{itemize}

\item[Figure 1.]{95 $\%$ CL limits on the $Z'$ mass as a function of the HERA
integrated luminosity arising from the ratio $R$. The solid(dashed) curve
corresponds to the GJS(BKN) model while the dashdot(dotted) curve is for the
LRM with $\kappa$=2(1) assuming $SU(2)_R$ breaking via isodoublet scalars.}
\item[Figure 2.]{95 $\%$ CL search limits for the $Z'$ in the the GJS model as
a function of the parameter $s_{\phi}$ at the Tevatron for several different
integrated luminosities assuming current electron and muon effeiciencies. From
top to bottom the first four curves are for 1000, 400, 100, and 25 $pb^{-1}$,
while the bottom dotted curve represents the current limits.}
\item[Figure 3.]{Same as Fig.2 but for the $Z'$ of the BKN model as a function
of the integrated luminosity.}
\item[Figure 4.]{(a) Same as Fig.2 but for the $W'$ in the LRM as a function of
the parameter $\kappa$. The limits obtainable $indirectly$ on the $Z'$ of the
LRM using the results of (a) and the $W'$,$Z'$ mass relationship for $SU(2)_R$
breaking via (b)doublets or (c)triplets of Higgs scalars. (d)The ratio of the
$Z'$ and $W'$ masses in the LRM as a function of $\kappa$ assuming doublet
(solid curve) or triplet(dashdot curve) breaking of $SU(2)_R$ used in
obtaining Figs.4b and 4c from Fig.4a. (e)The direct $Z'$ search limit at the
Tevatron for the LRM as a function of the integrated luminosity assuming that
$\kappa$=1.}
\end{itemize}

\end{document}